**TITLE PAGE**

Article type: Research

Manuscript title: Mobile Text Entry Behaviour in Lab and In-the-Wild studies: Is it different?

Names and affiliations of all contributing authors:

Andreas Komninos, University of Patras
Kyriakos Katsaris, Hellenic Open University
Emma Nicol, University of Strathclyde
Mark D. Dunlop, University of Strathclyde
John Garofalakis, University of Patras

Full address for correspondence, including telephone and fax number and email address
Andreas Komninos
Computer Engineering & Informatics Department
University of Patras
Rio 26504
Greece
+30 2610 997525
akomninos@ceid.upatras.gr

Acknowledgements:
This work was partly funded by project EP/K024647/1 (EPSRC UK)

This work was conducted in 2018 in partial fulfilment of the requirements for Katsaris' MSc thesis, available here https://apothesis.eap.gr/handle/repo/34050 (Greek only), using data, software and tools from the EP/K024647/1 (EPSRC UK) project https://pureportal.strath.ac.uk/en/projects/empirical-investigation-user-centred-development-of-touch-screen-

Manuscript prepared in June 2018

# Mobile Text Entry Behaviour in Lab and In-the-Wild studies: Is it different?

**AUTHOR INFORMATION**

Andreas Komninos[a], Kyriakos Katsaris[b], Emma Nicol[c], Mark Dunlop[c], John Garofalakis[c]

[a]*Computer Engineering and Informatics Department, University of Patras, Greece*

[b]*School of Positive Sciences, Hellenic Open University, Greece*

[c]*Computer and Information Science Department, University of Strathclyde, UK*

**ABSTRACT**

Text entry in smartphones remains a critical element of mobile HCI. It has been widely studied in lab settings, using primarily transcription tasks, and to a far lesser extent through in-the-wild (field) experiments. So far it remains unknown how well user behaviour during lab transcription tasks approximates real use. In this paper, we present a study that provides evidence that lab text entry behaviour is clearly distinguishable from real world use. Using machine learning techniques, we show that it is possible to accurately identify the type of study in which text entry sessions took place. The implications of our findings relate to the design of future studies in text entry, aiming to support input with virtual smartphone keyboards.

**Research Highlights**

- Even though it can seem intuitive that mobile text entry differs behaviour in lab and field settings, we demonstrate that these differences extend to the level where every text entry session can be accurately classified, no matter the context of the in-the-wild typing.
- We propose new low-level metrics to model user text entry behaviour, allowing the direct comparison of data from lab and field study settings
- We present a new way of analyzing and comparing text entry metrics across lab and field studies using machine learning techniques
- We highlight important future research directions for text entry research, both in devising new methodological tools for evaluating text entry in the lab, and for developing advanced text entry support in virtual keyboards.
- We openly share our methodology, data and software for the whole text entry community

## INTRODUCTION

Mobile text entry takes place primarily via virtual (touchscreen) keyboards. Text entry remains a critical element of the interaction with mobile applications and the overall user experience of smartphones and tablets. Most literature on the subject targets the improvement of text entry speed and accuracy, through the design of different keyboard layouts, underlying error correction mechanisms and interaction styles (e.g. swiping). It is also true that most literature relates to findings from studies in lab-based experiments, primarily via copy (transcription) tasks, were participants are shown sentences and are asked to type these sentences in. A smaller body of literature includes studies in-the-wild, or approximations of real world use in a lab (e.g. typing while walking). Although the contribution of such studies is significant, we were able to find only one study in which the users' text entry behaviour in the lab and in the real world is directly compared, thereby raising the question of how well lab-based use matches real world use, and concerns for the generalizability and of such studies.

Motivated by this, we present in this paper the results of an attempt to investigate whether lab-based text entry differs from real world behaviour. For this, we used data from one lab-based experiment and two in-the-wild studies, to train machine learning classifiers and assess their ability to distinguish between the two types of study setting.

## RELATED WORK

The debate over the utility of lab vs. field based studies in HCI has been ongoing for more than a decade. In a review of past practice, (Kjeldskov & Skov, 2014) highlight that the mobile HCI community is still divided regarding the utility of both types of study. According to their recommendations, field studies are necessary so that we can understand actual user behaviour in realistic settings. Thus, according to the authors, the question emerging from the review of mobile HCI literature is not whether we should carry out field studies, but when, and how.

In-the-wild studies of mobile HCI applications are increasingly common, however, one area in which such studies are extremely rare, is text entry. While it is one of the most common functions performed on smartphones (and, increasingly, wearables such as smart watches), text entry has seldom been evaluated outside the lab. One contributing factor for this, is that it was almost impossible, without full access to the operating system source code, to replace the default input methods (e.g. virtual keyboards) on a device with a novel design. This situation changed in 2011, when it became possible for developers to construct new input methods for the Android OS (since version 3.0), which could fully replace the default keyboard of the device, and to distribute these as downloadable applications, suitable for general everyday use.

Although one would expect that the text entry research community would quickly embrace this capability and begin to shift its efforts towards studying user input behaviour in the field, it seems from our review of post-2011 literature, that the uptake was quite slow. One possible explanation is that planning and conducting a study that asks users to replace a core part of their mobile experience (i.e. the keyboard) with a research version that doesn't necessarily include all the supportive features that commercial keyboards have, is perhaps an overly difficult task for researchers. It requires significant investment in recruiting, handling privacy issues, installing, monitoring use and enumerating participants. The convenience of lab studies is their relatively cheap and quick turnaround of results, and for this reason it is easy to see the appeal to researchers.

We reviewed existing literature using Google Scholar as the main search engine, since it has been shown that its coverage outperforms that of single repositories (Gehanno, Rollin, & Darmoni, 2013) (Harzing, 2014) (Martín-Martín, Orduna-Malea, & López-Cózar, 2018). We queried for key terms (e.g. "smartphone", "mobile", "text entry", "keyboard", "input"), selecting articles published on or after 2011 only, and focusing on smartphone rather than smartwatch input, since we wanted to gain a sense of the progress and state-of-the-art in smartphone input research during recent years.

Our review of literature focuses on the type of study and the metrics used therein, and not on the actual foci of the study, with the aim to establish the current standards of study practice in the mobile text entry community. Since our article is not a literature survey paper, we present here some representative examples for the various foci of research pertinent to text entry in smartphones. The result is a non-exhaustive but adequately representative literature search (Table 1), from which we reached two important conclusions.

### Type of studies in mobile text entry research

Firstly, with regard to the type of study, the overwhelming majority of user studies are lab-based and rely on transcription tasks (i.e. participants are being shown specific sentences on the device and are asked to type them back in). Although it is generally accepted that such types of study are the de-facto way of testing mobile text entry methods (Reyal, Zhai, & Kristensson, 2015), other studies such as (Clawson, Starner, Kohlsdorf, Quigley, & Gilliland, 2014)(Lin, Goldman, Price, Sears, & Jacko, 2007) show that real world conditions can significantly impact user text entry behaviour. Naturally this is does not retract the value and contribution of lab-based studies, however, some doubt might be placed on the ecologic validity of models and interfaces built on the experience and data from lab studies alone. Although it is generally accepted that user performance may improve with familiarity with a keyboard (Dalvi et al., 2016), most lab-based studies rely on single sessions of evaluation, as longitudinal, repeated-measure studies are expensive and take a long time to complete, without guarantee that the participants will reach "expert" levels of performance (Banovic, Rao, Saravanan, Dey, & Mankoff, 2017). Of the literature we reviewed, only two studies required participants to attend multiple sessions. A few studies go to some lengths to simulate real-world conditions (2 by having participants walk as they type, 1 typing while driving a car simulator). We were able to discover three studies that assesses typing performance in the wild. In (Henze, Rukzio, & Boll, 2012), text entry behaviour was recorded in the context of a mobile game that required users to enter words as they appeared on the screen. Various design modifications of the baseline keyboard were evaluated for speed and error rate performance using the game. In (Reyal et al., 2015), participants were asked to perform transcription tasks during their daily life, prompted regularly to perform the task by a software service running on their device. Hence this study is not a pure observation of real behaviour, but in essence, a replication of the lab procedures outside the strict confines of the lab. In (Buschek, Bisinger, & Alt, 2018), real world text entry behaviour in various apps is recorded by installing a replacement input method on the smartphones of users. The researchers claim this to be the first study that actually reports on real typing behaviour. The study offers a statistical account of users' performance (words-per-minute, WPM), hand posture, device orientation, use of autocorrection, finger slippage and text deletion. The researchers compare some of their aggregate findings to the empirical outcomes reported in a few lab study papers, but this comparison is in the form of general commentary, and not in any way a systematic or methodologically valid effort.

### Metrics and effect of task design in mobile text entry research

The second conclusion relates to the metrics primarily used in text entry research. Almost all studies rely on reporting text entry speed (words per minute) and error rates, with other proposed metrics, e.g. (Soukoreff & MacKenzie, 2003), remaining largely unused in mobile text entry research. Therefore no other user behaviour is examined, although researchers have identified that factors such as context-switching (i.e. shifting the focus of attention to check for input errors, or word completion suggestions) can impact user cognition and thus entry speed (Bi, Ouyang, & Zhai, 2014). Where other entry behaviour (e.g. finger slippage) is examined, this is mostly used to build touch models to improve input accuracy and performance by disambiguating touch intentions, e.g. (Weir, Pohl, Rogers, Vertanen, & Kristensson, 2014). The use of error rates as a performance metric is possible due to the nature of the transcription task: since the researchers pre-determine the user's intended input by specifying the text to be entered, they can measure their performance in terms of final submitted text accuracy, and monitor for omission, substitution, proximity and other common error types during input. The use of WPM as metric of performance is also dominant in lab studies, and even reported in some of the (few) field studies we were able to find. Early research such as (Salthouse, 1986) has shown that observed entry speed depends on both cognitive and motor skills, which can vary across individuals and, of course, can be trained and improved over time, e.g. see (Lelis-Torres, Ugrinowitsch, Apolinário-Souza, Benda, & Lage, 2017). Thus, input speed metrics such as WPM constitute a high-level metric, which combines subjects' cognitive and motor skill performance. In (A. S. Arif & Stuerzlinger, 2009) evidence is found that the way human errors are handled during text entry has an effect on the common text entry metrics, therefore using these metrics without proper account for error correction strategies may yield misleading results.

As stated previously, the text entry community acknowledges the existence of a learning effect (shown in repeated WPM measurements across multiple sessions), but to our knowledge, there has been no literature so far to investigate the cause of this effect (i.e. how much of it is attributable to cognitive improvements vs. motor skill improvement). Work such as (Jokinen et al., 2017)(Das & Stuerzlinger, 2008) shows that cognitive factors relating to recall of the keyboard layout definitely play a significant role in observed text entry speed. Still, it remains unknown whether improvements in repeated measure studies are due to increased familiarity with the keyboard, or simply increased familiarity with the laboratory procedure (e.g. knowing what to expect, feeling less stressed, being able to recall phrases from previous sessions, getting better at memorizing larger chunks of sentences and hence not having to shift attention focus during typing etc). A demonstration of the effect that task design can have on the observed participant performance can be seen in (Soboczenski, Cairns, & Cox, 2013), where it was found that decreasing the legibility of presented transcription phrases improved the participants' capacity to memorize the content (since they had to focus on it more heavily) and thus improve the overall input speed, as the participants had less need for context-switching between reading and typing.

| Authors | Setting | Focus | Metrics | Population | Sample |
|---|---|---|---|---|---|
| (Ahmed Sabbir Arif, Sylla, & Mazalek, 2016) | Lab | Colour coding of autocorrection | Speed, error rates | Children | 26 |
| (Azenkot & Zhai, 2012) | Lab | Touch model | Speed, error rates | Young adults | 32 |
| (Banovic et al., 2017) | Lab | Error correction behaviour | Speed | Young adults | 20+8 |
| (Bi et al., 2014) | Lab | Word completion algorithms | Keystroke saving | Young adults | 15 |
| (Bi & Zhai, 2016) | Lab | Gesture keyboard layouts | Speed, error rates | Young adults | 24 |
| (Buschek et al., 2018) | Field | Virtual QWERTY | Speed, finger slippage, autocorrect use, error correction | Young adults | 30 |
| (Clawson et al., 2014) | Lab - simulated environment | Physical QWERTY | Speed, error rates | Young adults | 36 |
| (Dalvi et al., 2016) | Lab - longitudinal | Non-english tap & gesture keyboards | Speed, error rates | Children | 153 |
| (Dunlop, Komninos, Nicol, & Hamiliton, 2014) | Lab | Physical gestures for virtual QWERTY | Speed, error rates | Young & older adults | 123+23 |
| (Gkoumas, Komninos, & Garofalakis, 2016) | Lab | Visibly adaptive virtual layout | Speed, error rates | Young adults | 20 |
| (Henze et al., 2012) | Field | Virtual QWERTY | Speed, error rates, finger slippage | Unknown | ~73,000 |
| (Hakoda, Shizuki, & Tanaka, 2016) | Lab - longitudinal | Virtual QWERTY | Speed, error rates | Young adults | 12 |
| (Jo & Cho, 2015) | Lab | Alternative virtual layout | Speed, error rates | Young adults | 10 |
| (Jokinen et al., 2017) | Lab – longitudinal | Alternative virtual layout | Gaze, visual search time | Young & older adults | 33+10 |
| (Kai, Tsukamoto, & Iio, 2017) | Unclear | Japanese virtual keyboard | Speed, error rates | Young adults | 21 |
| (Liu, Dillon, & Zhang, 2017) | Lab | Physical QWERTY, pen input | Speed | Older adults | 18 |
| (Ljubic, 2016) | Lab | Tilt-based input | Speed, error rates | Young adults | 20 |
| (Munger et al., 2014) | Lab - simulated environment | Virtual QWERTY | Task completion time | Young & older adults | 24 |
| (Pritom, Mahmud, Ahmed, Hasan, & Khan, 2015) | Lab | Alternative virtual layout | Speed | Young adults | 10 |
| (Reyal et al., 2015) | Lab, field | Virtual QWERTY, gesture QWERTY | Speed, error rates | Young adults | 12+12 |
| (Ruan, Wobbrock, Liou, Ng, & Landay, 2018) | Lab | English & Mandarin virtual QWERTY / pinyin | Speed, error rates | Young adults | 24+24 |
| (Ryu et al., 2016) | Lab | Visibly adaptive virtual layout | Speed, error rates | Older adults | 30 |
| (Sarcar et al., 2017) | Lab | Alternative virtual layout | Speed, error rates | Older adults | 2 |
| (A. L. Smith & Chaparro, 2015) | Lab | Physical QWERTY, virtual QWERTY, tracing, handwriting, and voice | Speed, error rates | Young & older adults | 25+25 |
| (B. A. Smith, Bi, & Zhai, 2015) | Lab | Gesture QWERTY | Speed, error rates | Young adults | 14 |
| Tani & Yamada (Tani & Yamada, 2015) | Lab | Touch model | Tapping errors | Young adults | 10 |
| Weir et al. (Weir et al., 2014) | Lab - simulated environment | Touch & language models | Speed, error rates | Young adults | 10+16 |
| Yeo et al. (Yeo, Phang, Castellucci, Kristensson, & Quigley, 2017) | Lab | Virtual QWERTY, gesture QWERTY, physical gestures for virtual QWERTY | Speed, error rates | Young adults | 12 |

**Table 1. Overview of the characteristics of reviewed smartphone text entry studies (2011-on).**

The commentary above is supported by Table 1 which summarizes the recent related literature we reviewed. As a result, we are left wondering whether text entry differs between the typical transcription-task based lab study and real-world settings, beyond the obvious error-rate and input speed metrics, which are the composite result of a number of underlying factors.

The rest of the paper describes our analysis of data from a laboratory study using a simple virtual QWERTY smartphone keyboard, and conducted in the typical manner described above, and the data collected from in-the-wild use of the same keyboard.

## STUDY CONSIDERATIONS

Our literature review indicated that text entry research has hinged for many years on the assumption that the typical lab-based study offers adequate external validity to approximate real world use. It is understandable that it wasn't until a few years ago that it became technically possible for researchers outside device manufacturing and operating system companies, to develop input methods that could be used in-the-wild as a replacement for standard virtual keyboards. However, we were surprised to discover that even to this date, the text entry community hasn't attempted to examine real world text entry, against text entry in the lab. Therefore, our main aim was to investigate whether any differences could be found. If differences were discernible, then this would make a compelling argument towards either a shift for more field-based evaluations in the future, or the evolution (or even abandonment) of the traditional transcription-task lab study, and a push towards the design of new lab-based methods that can offer better external validity to our research. To do this, our research needed to make the following considerations, described next.

### Selecting a direct replacement input method with logging ability

First, we needed to develop an input method (IME) which can replace the standard input method on Android devices (the most popular OS) and that is able to log metrics of users text entry behaviour. With such a tool, we could perform both a lab-based study and an in-the-wild collection of text entry data, using the same input method and therefore be able to compare the collected data. For this purpose, we used the open-source QWERTY MaxieKeyboard implementation available on GitHub (Komninos, Nicol, & Dunlop, 2015). This keyboard allows the logging of various text entry parameters for research and has the advantage that logging is performed by the keyboard software itself, thereby allowing its use as a complete replacement for the Android OS OEM-installed keyboards. The GitHub package also comes with server-side code, including an API and database structure, so that collected data from participants' smartphones can be remotely uploaded to a server for storage and later analysis.

### Selecting metrics for comparison

Secondly, we needed to consider the type of metrics that could be used to compare lab and field study behaviour. When considering this problem, the obvious answer might be to use the traditional WPM and error rate metrics, but a comparison with these two metrics would be inappropriate, and we elucidate the reason with a simple example. Regarding WPM, we know from (Salthouse, 1986) that input speed is dependent on cognitive and motor ability. In a lab setting, the user's attention can be reasonably considered to focus exclusively at the task at hand, without external disruptions. However, in the field, this assumption cannot be made. If a user is typing a text message while standing in a bus, their attention is divided between the text entry activity, paying attention to other passengers, trying not to fall, looking out of the window to check how close they are to their bus stop, and maybe dismissing incoming notifications on their mobile. Because it is impossible to fully capture the user's context while performing a real world typing activity, we are unable to control, or account for influencing factors, and thus it would be methodologically wrong to try to compare a high-level metric of the subjects' performance (such as WPM) as measured in the field, with their performance in a lab, where all external factors are tightly controlled. As such, a comparison using WPM is inappropriate (and hence our earlier comment about the comparison presented in Buschek et al., 2018).

Additionally, regarding error rates, MaxieKeyboard is able to detect spelling mistakes as a user completes typing a word and log this information. In theory we could compare this information with the same data in the lab, but in reality, this metric is also problematic. Real world text entry is messy and full of colloquialism, grammatically and syntactically incorrect use of language, acronyms and new words that are invented, used and forgotten as online culture emerges and fades, e.g. see (Stieglitz & Dang-Xuan, 2013). Therefore, logging such purposely “wrong” input as errors does not fit particularly well into a fair comparison with lab-based input during transcription tasks, where the user’s intent for input is strictly dictated by the researchers. Another reason for not using this metric is quite simply the different nature of the task during field entry: Since there is no prescribed text that the user has to input, the user is quite free to change their mind mid-word regarding what they might like to input. Therefore, the deletion of a character or a sequential deletion of multiple characters or words, does not necessarily indicate the occurrence, detection and correction of an input error, as it clearly would in a transcription lab task.

To overcome these problems, it is clear that we need to record lower-level text entry metrics that can be directly compared in any study setting. The time spent in each of the cognitive and motor phases of input is reported as non-Fitts and Fitts time in (Das & Stuerzlinger, 2008). In the case of lab experiments, non-Fitts time includes time spent reading and memorizing parts of the phrase to be transcribed, while in the field, this time includes reflecting on what to actually write, along with dealing with external interruptions. In both types of experiment, some non-Fitts time is required to visually seek the appropriate key to press, although in the case of expert users familiar with the keyboard layout, this can be considered to be minimal. Since it is shown in (Salthouse, 1986) that there exists the concept of minimum commitment and that users memorize sentences in “chunks”, it is reasonable to think that “pausing” behaviour due to non-Fitts time is more likely to exist between the typing of individual words, hence WPM is most likely affected by such pauses. To obtain a more equitable comparison between the two settings, it is therefore more appropriate to use inter-key time as a measure of typing speed that largely excludes the time spent during a user’s context switch between actual typing and mentally focusing on the requirements of the task at hand (reading, or reflecting).

A further metric not previously considered in literature but actually closely related to the motor skills that make up Fitts’ time during typing, is the duration of a keypress (i.e. the elapsed time between touch-down and touch-up events on a virtual keyboard). A longer keypress obviously adds to the overall Fitts’ time and therefore also affects input speed. Recent evidence (Ciman et al. 2015) also exists that this metric might be closely linked to a user’s emotional state (i.e. longer duration associated with increased stress) and thus affecting their performance during non-Fitts time phases.

Keypress duration is also likely associated with finger slippage along the horizontal and vertical axes, which has been examined in the past and found to be a significant contributor to erroneous input [ref]. As we explained previously, a comparison using logged spelling errors or character deletions would be inappropriate. However, finger slippage is an objective and directly comparable error-causing behaviour, and thus we could use this metric instead.

Therefore, instead of using a calculated WPM metric from the data logged by MaxieKeyboard, we use the keyboard’s ability to record low-level data for every keystroke. Keystrokes are grouped into “sessions” which begin at the moment the keyboard is shown on the screen, and end when the keyboard is retracted. For each keystroke thus, the following data is recorded:

- **X-axis difference** between the touch down and touch up coordinates of each keypress (pixels)
- **Y-axis difference** between the touch down and touch up coordinates of each keypress (pixels)
- **Inter-key time** as the time elapsed since the previous keypress (milliseconds)
- **Duration** of keypresses i.e. how long the finger made contact with the screen (milliseconds)

The first two are representative of behaviour that is likely to lead to input errors, while the latter are representative of behaviour that might affect input speed. The last two could be added to estimate entry speed, if the number of touch events recorded resulted in actual entered characters. However this is not the case with mobile keyboards, as may touch events correspond to punctuation, shift, backspace, language change, symbol mode change or other non-entry events. MaxieKeyboard can also record the keycode and corresponding character, but we disabled this feature to preserve user privacy and protect sensitive information in the field experiments (we recorded only non-sensitive characters, such as punctuation marks and symbols for purposes that we explain next).

### Selecting a data analysis method

Finally, we needed to consider the method that could be used to compare data across two entirely different set-ups. In the lab, we have a controlled and fixed amount of text entry activities per user, offering a finite number of data points, all under the same usage scenario (typically approximating the composition of a short message). In the wild, the number of activities that can be performed per participant are unforeseeable, their input length varying also with the nature of the usage context (e.g. typing an SMS, a comment in social networks, entering a URL or email address, typing a telephone number or their credit card details). To compare between the two, using typical statistical inference methods (e.g. rANOVA or Kruskal-Wallis tests) as per common practice in the literature might not be suitable, not just because of the uneven volume of data and number of data points, but because it is uncertain how data from the field study might be selected, filtered and then tested to examine whether the assumptions for the various types of statistical tests hold. Other methods such as mixed models require a number of assumptions which are simply impossible to make for field data (or, to meet assumptions, exclusion criteria need to apply to the data and it is impossible, without context, to know what constitutes a valid case and not an outlier). Finally, methods which would seem appropriate, like binomial logistic regression, assume independence of observations, therefore requiring that each participant's data is averaged before fed into the model, thereby discarding a lot of information which might otherwise be very useful.

Statistical methods have the advantage of being able to identify differences but also of being able to attribute causality to the factors contributing to those differences, by uncovering the degree to which a factor contributes to the variance of data. In a tightly controlled laboratory environment, the strong internal validity of the experiment guarantees that we are able to use statistics to attribute causality with confidence. However, in our case, this advantage would be largely negated, due to the nature of the field data. In essence, because it is impossible to have a full understanding of the context under text entry behaviour data is generated in the field, we could determine that a certain factor (e.g. the keypress duration as the time elapsed between key-down and key-up events) has a certain effect, but the actual reason why this effect is caused would remain unknown (for our duration example, it stress, hand and device posture, hand strength or some other confounding variable that we could not capture? Is it even the actual participant and not a temporary user of the device affecting the results?).

A more recent trend in data analyses is the use of machine learning tools. Although such approaches are quite novel in the analysis of mobile HCI data, they are gaining ground as a tool in various mobile HCI contexts (Mayer, Le, & Henze, 2017), including text entry (Ghosh, Ganguly, Mitra, & De, 2017). In the context of our objective, a traditional approach would be to examine data from two labelled categories (i.e. observed from a lab or field study), with an aim to determine whether we can reject or uphold the null hypothesis (i.e. that the means of a given metric that characterizes input behaviour in these two conditions are the same), or to attempt to build a multi-factor model that describes the observed data, and show that the weights for each factor are different in the two categories of datasets. In machine-learning terminology, this process can be considered equivalent to binomial classification of data. Just as we might infer, for example, that user input behaviour in the lab and field are different A machine learning algorithm can be trained to detect whether input data belongs to one of two classes ("lab" or "field"). A classification accuracy of 50% in this case offers no better result than a random assignment. As the accuracy of the classifier rises, so can the confidence in claiming that there exist enough discernible differences in the data, so much so that an automated tool can successfully spot them. Machine learning tools have the advantage of being agnostic of the data and their inherent attributes, and require no assumptions as do statistical tests (e.g. sphericity, distribution etc). They also result in a predictive model that can be applied to a range of validation data to test for accuracy, or can be later used to predict outcomes based on input samples (e.g. quite easy to incorporate in a mobile app, that monitors the user's current context, either on-device or off-loading computation to the cloud). On the negative side, machine learning algorithms are largely "black boxes" in that they can provide a result, but the effect of each factor in the contributing result remains hidden. Given however the low internal validity of field studies, the subject of causality is a non-factor in the consideration of an analysis method. We thus selected to use two common predictive modelling techniques, using Neural Networks and Support Vector Machines.

## DATASETS AND STUDY SETTINGS

Three data collection processes were performed with MaxieKeyboard. Two of them were in-the-wild studies and one experiment was a lab-based study. We did not use any of the keyboard's advanced features, except the suggestion bar, resulting in a plain QWERTY implementation. In all studies we recruited participants who did not use gestural text-entry techniques (e.g. swiping) in their daily life.

For the lab study, proficient English-speaking participants (under & post-graduate computer science students at the University of Patras) performed a simple copy task, where they were shown 15 randomly selected phrases between 50-80 characters long from the Enron memorable phrase set, removing all punctuation. Participants entered text without time restriction and the sole instruction to proceed as quickly as comfortable to them, and as accurately as possible. All participants reported as highly familiar with using QWERTY virtual keyboards and considered themselves "expert" smartphone typists. Participants were given 10 minutes of free practice with the keyboard and experiment software, which consisted of a simple interface that presented a phrase and a text entry field directly below the phrase. The experiment was performed on a Nexus 5 smartphone.

This experiment setup is quite typical in many text entry studies, in terms of the number of phrases used for transcription tasks and volume of performed tasks (e.g. see (A. L. Smith & Chaparro, 2015), (Walker, Li, Vertanen, & Kuhl, 2017)), sample size and use of convenience sampling (Vertanen, Fletcher, Gaines, Gould, & Kristensson, 2018), thus replicating the study setup as commonly encountered in literature.

For the two field studies, there are no established practices in the text entry community regarding how collection should be made, sample sizes etc. In (Buschek et al., 2018), data was collected from 30 participants aged 18-33 over a period of 3 weeks and all data entered in any context and application was included in their analysis. Our studies included a preliminary collection trial with 3 computer science student participants, and a further collection study with 12 graduate (non-student) participants. Both studies spanned a period of 28 days and participants were from two different universities, unrelated to the lab study population. In both field studies, MaxieKeyboard was installed on the participants' own device, replacing the OEM keyboard. As we will explain next, we did not include all recorded data in our analysis, since our aim was to be able to directly compare lab and field text entry. Details of the studies and the participants are shown in Table 2.

| Dataset | Use context | Valid sessions | Collection period | Users | Ages |
|---|---|---|---|---|---|
| DS1-F | Field | 1243 | 28 days | 12 (6f) | 25-35 |
| DS2-F | Field | 542 | 28 days | 3 (1f) | 20-25 |
| DS3-L | Lab | 298 | N/A | 20 | 18-35 |

**Table 2. Details of data collection studies and participants. Valid sessions are those with characteristics akin to lab study text entry.**

### Field data cleansing

Further from individual keystroke metrics, for text entry sessions (text entry during the appearance of the keyboard on the screen), MaxieKeyboard records the start and end time of the session, as well as the name of the application that invoked the keyboard. To construct our final field study datasets, we filtered the field data so as to keep only those text entry sessions that would be most directly comparable to the data obtained in the lab study. Since the lab study phrase corpus and task emulates the input of text messages, we kept only those field entry sessions which were generated in messaging and emailing applications (e.g. SMS, Facebook Messenger, WhatsApp, Gmail). From those, we also included only sessions with 10-160 keystrokes and further filtered to exclude sessions that contained punctuation other than periods and commas, since other special characters required a shift of the keyboard to symbol mode and would thereby artificially alter inter-key time (the user has to search for the symbols first). This data-cleansing process allowed us to generate data that is directly comparable with the data from the lab experiment. Finally, for each dataset, we averaged out the data across each session, thereby constructing our final datasets consisting of averaged session metrics, as per Table 2.

## CLASSIFIER TRAINING AND VALIDATION DATA

We carried out two rounds of experiments (A & B) in which the training set for the classifier consisted of all the data from the lab experiment (298 sessions) and an equal number of random data from each of the field experiment sets. For each experiment we constructed 15 such combinations of training sets, choosing for each set randomly from the respective field data pool. For validation we used the entire DS2-F dataset, and a randomly selected subset from DS1-F of equal size (selected only once). For each of the sessions in the training and validation sets, we calculated the averages of the text entry behaviour feature values, clearing out events that were extreme outliers (± 2 standard deviations from the mean). The processed data was used then to train the classifiers under a range of parameters, and consequently, to perform classification.

Before proceeding, a note on our parameter selection approach here is required. In standard machine learning research practice, algorithm performance is evaluated using three separate datasets: First, a set is used to optimize parameters (often using an exhaustive grid search or genetic algorithm), effectively training and validating the algorithm's performance over many iterations, to select the parameter combination that provides the best performance on that dataset. Then, a second independent set is used to train the algorithm and build a model, and finally, the performance of the model is assessed using a third, unseen dataset. This approach ensures that the model is not overfitted to the training data, and thus guarantees the objective performance assessment of the algorithm and its general performance for any randomly selected input. In our case, we are not interested in proving that a given algorithm and its resulting model under a fixed training set and parameters outperforms another, but instead, we are interested in finding what the best possible performance of an algorithm can be, by exploring a range of parameters and assessing that performance over multiple datasets. In essence, we perform step 1 of the machine learning research practice process, since a good accuracy achieved in that step is enough to prove our hypothesis (i.e. that lab and field text entry behaviour is clearly distinguishable). The fact that we use a multitude of random datasets for training guarantees the objectivity of the outcome, since it is impossible to assume that the result is the product of overfitting the model (via biasing parameter selection) to a single dataset.

Therefore, a total of 30 training and 2 validation sets were used to feed and test the performance of two types of classifiers (Neural Network and SVM), and we report the results of our analysis next. The composition of the training and validation sets is shown in Table 3 and the descriptive statistics of each of the validation sets and the DS3-L set used for training is shown in Table 4. In our analysis, we measured both prediction (classification) accuracy and execution time, the latter for the purposes of assessing the feasibility of on-device or cloud-based execution of the classifiers in future work.

| | Training set (50% + 50%) | | Training set pool size | Final Training set size | Validn. set | Valid. set size |
|---|---|---|---|---|---|---|
| **A** | DS1-F | DS3-L | 1541 | 596 | DS2-F | 542 |
| **B** | DS2-F | DS3-L | 840 | 596 | DS1-F | 542 |

**Table 3. Training & validation sets used in experiments A & B**

| | Mean values (std. dev.) | | | |
|---|---|---|---|---|
| **Dataset** | **X-diff** | **Y-diff** | **Interkey** | **Duration** |
| DS1-F (V) | 0.565 (2.944) | 1.455 (1.795) | 411.557 (384.427) | 133.646 (39.699) |
| DS2-F (V) | -1.100 (1.441) | 0.145 (2.162) | 527.625 (397.299) | 97.034 (20.419) |
| DS3-L | -0.040 (1.150) | -0.514 (1.298) | 432.905 (239.384) | 165.280 (22.431) |

**Table 4. Descriptives of validation datasets and DS3-L**

## CLASSIFICATION RESULTS & ANALYSIS

All experiments were conducted using a simple workstation computer (Intel Core i5 2.40GHz, 4GB RAM, Windows 7), on which all unnecessary software and services were suspended and the computer was disconnected from the network. For the calculation of results, we used the RapidMiner Studio software. We measured the outcome (classification accuracy) and running time, as reported by RapidMiner. In the analyses below, where statistical significance tests are reported, the choice of test is based on the distribution of data (Shapiro-Wilk normality test).

### Using Neural Networks

RapidMiner allows the parametrization of neural networks (NN) with a range of options, for which we accepted the recommended defaults as a starting point. Namely, our parameters were as follows: Hidden layers (not specified, resulting in a sigmoid HL type, of size (n_attributes+n_classes)/2+1), training cycles (10,000), learning rate (0.01), shuffle (yes), momentum (0.3), decay (yes), normalize (yes). Of greatest interest here are the learning rate and momentum parameters, which critically define the behaviour of the NN and the subsequent results, either through the possibility of overfitting the model (learning rate) or being trapped in local minima (momentum). Therefore, to obtain the best possible results, we decided to experiment with setting these two parameters as follows. First, we accepted the default learning rate and ran each experiment subset (15 runs each), varying the momentum parameter with values between [0.3, 0.8] in steps of 0.1 (6 values). From these results, we kept the best performing momentum value in terms of prediction accuracy and execution time. Then, we repeated the execution of the 15 subsets in each experiment for varying values of learning rate, using values between [0.001 – 0.01] in steps of 0.001 (10 values). Based on the accuracy of results, we determined the optimal NN performance, which is reported next.

#### *Experiments A & B - selecting a momentum value*

For experiment A, a surprisingly good accuracy of 100% (sd=0.00%) was achieved in all but one settings of momentum ([0.3,0.7]) (Figure 1). A Friedman test revealed a statistically significant difference between the accuracy at each momentum setting ($\chi^2_{(5)}$=21.842, p=0.001). However, multiple post-hoc Bonferroni corrected pairwise Wilcoxon signed-rank tests, resulting in a significance level set at $p \leq 0.0033$, reveal that there is no statistically significant difference in the accuracy between any two momentum settings. Accuracy performance can hence be considered equal across all momentum settings.

In terms of execution time (Figure 2), a Friedman test revealed a statistically significant difference between the accuracy at each momentum setting ($\chi^2_{(5)}$=37.628, p=0.000). Post-hoc Bonferroni corrected pairwise Wilcoxon signed-rank tests, resulting in a significance level set at $p \leq 0.003$, reveal that there is a statistically significant difference in execution time for between momentum values 0.3 & 0.8 ($\Delta_{Mean}$ = 0.93s, Z=-3.276, p=0.001), 0.4 & 0.8 ($\Delta_{Mean}$=0.80s, Z=-3.207, p=0.001) and 0.5 & 0.8 ($\Delta_{Mean}$ = 0.87s, Z=-3.357, p=0.001). Based on these results, we selected to proceed with momentum value 0.4, since its nominal mean execution time (M=6.13s, stdev=0.34s) is only 0.80s greater than that of momentum value 0.8, but it offers a 100% nominal average accuracy and is also the fastest between those value options that offer the best nominal average accuracy (i.e. 0.3, 0.4 & 0.5 which all have M=100.00%, stdev=0.00%).

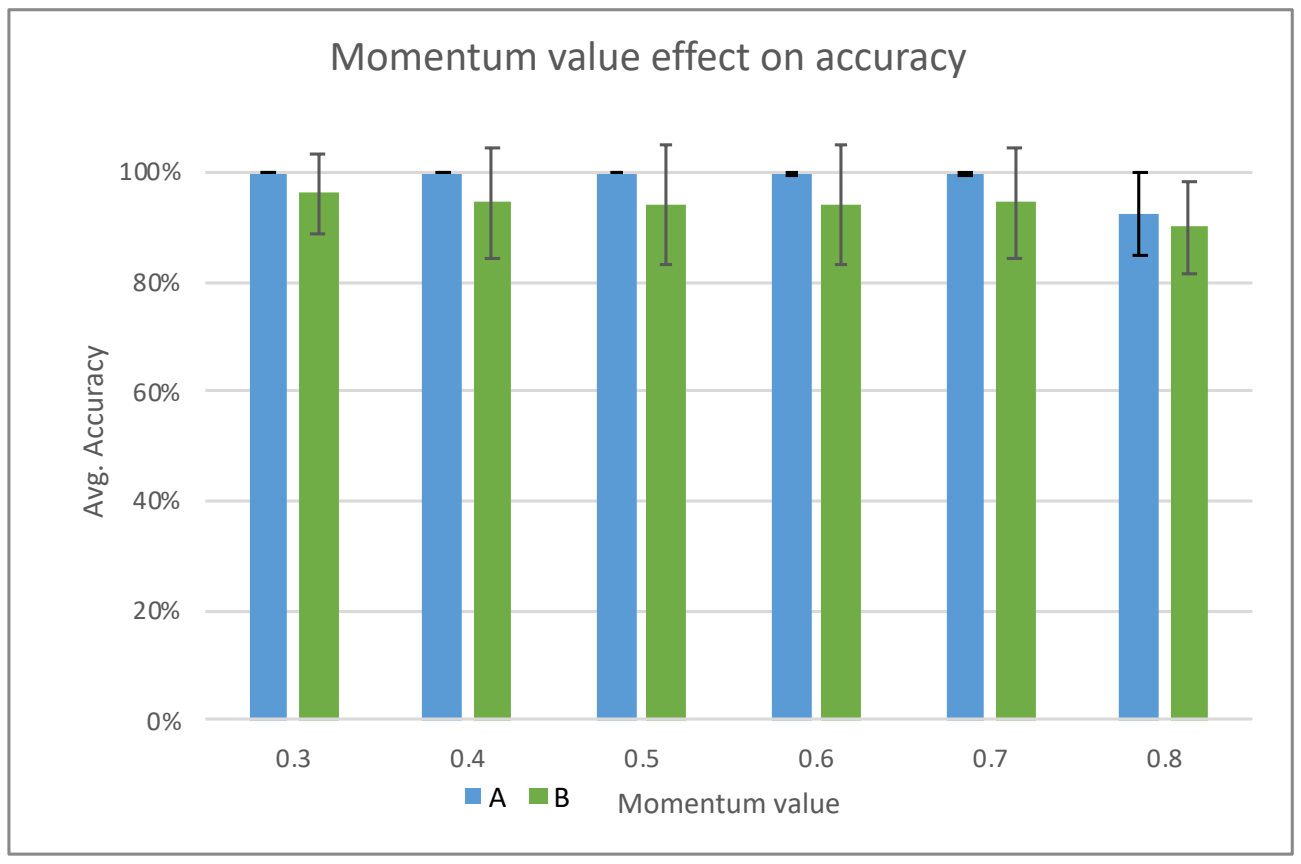

**Figure 1. Effect of momentum value parameter on classification accuracy with NNs (LR=0.01). Error bars at 95%c.i.**

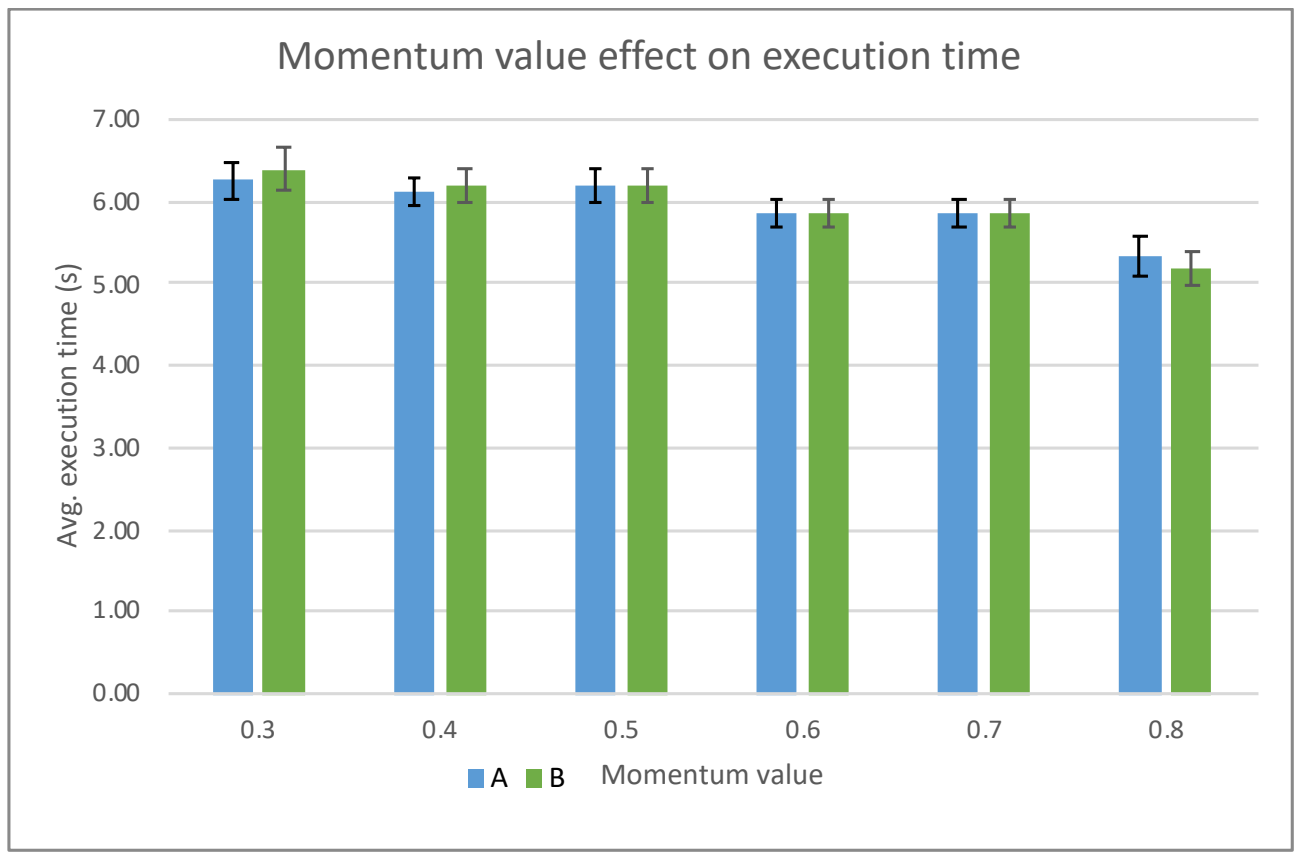

**Figure 2. Effect of momentum value parameter on execution time with NNs (LR=0.01). Error bars at 95%c.i.**

Following the same process for experiment B, classification accuracy varied more but remained firmly above 90% for all momentum values (Figure 1). A Friedman test revealed a statistically significant difference between the accuracy at each momentum setting ($\chi^2_{(5)}$=33.595, p=0.001). Multiple post-hoc Bonferroni corrected pairwise Wilcoxon signed-rank tests, resulting in a significance level set at $p \leq 0.0033$, reveal that there exists a statistically significant difference in the accuracy only between values 0.3 (m=96.12%, stdev=14.46%) and 0.8 (m=90.04%, stdev=16.49%) with Z=-2.938, p=0.003.

In terms of execution time, we note a linear reduction trend as the value increases (Figure 2). A Friedman test revealed that the observed differences are statistically significant ($\chi^2_{(5)}$=45.066, p=0.000). Post-hoc Bonferroni corrected pairwise Wilcoxon signed-rank tests, resulting in a significance level set at $p \leq 0.003$, showed statistically significant differences exist in the combinations shown in Table 5. From these results we selected to proceed with a momentum value of 0.3, as it offers an advantageous and statistically significant difference of both accuracy and execution time between itself and value 0.8.

| LR value pairs | Execution time (s) Mean | Std. Dev. | Test statistics Z | p |
|---|---|---|---|---|
| 0.001 - 0.008 | 7.40 - 6.20 | 0.49 - 0.40 | -3.286 | 0.001 |
| 0.001 - 0.009 | 7.40 - 6.20 | 0.49 - 0.40 | -3.286 | 0.001 |
| 0.001 - 0.010 | 7.40 - 6.20 | 0.49 - 0.40 | -3.286 | 0.001 |
| 0.002 - 0.005 | 6.93 - 6.20 | 0.25 - 0.40 | -3.317 | 0.001 |
| 0.002 - 0.008 | 6.93 - 6.20 | 0.25 - 0.40 | -3.317 | 0.001 |
| 0.002 - 0.009 | 6.93 - 6.20 | 0.25 - 0.40 | -3.317 | 0.001 |
| 0.002 - 0.010 | 6.93 - 6.20 | 0.25 - 0.40 | -3.317 | 0.001 |

**Table 5. Exp A. statistically significant differences of NN execution time for learning rate values**

*Experiments A&B – Learning rates and results with NNs*

For experiment A, we proceeded to repeat the NN classification process using multiple values for the learning rate. In each case, and for all learning rate values, the result was a 100% accurate classification (m=100%, stdev=0.00%) (Figure 3), therefore the modification of the learning rate value does not affect the classifier's ability to distinguish between lab and field typing behaviour.

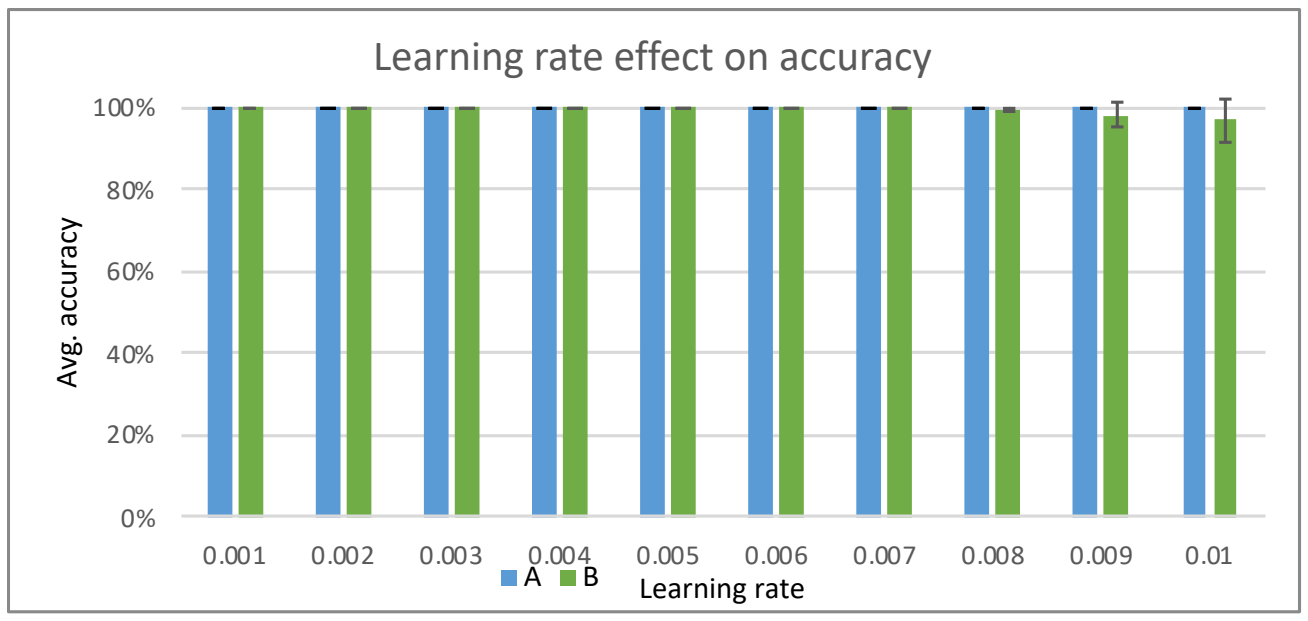

**Figure 3. Effect of learning rate value parameter on NN accuracy (momentum=0.3). Error bars at 95%c.i.**

In measuring execution time (Figure 4), a Friedman test revealed a statistically significant difference between the accuracy at each learning rate setting ($\chi^2_{(9)}$=82.219, p=0.000). Post-hoc Bonferroni corrected pairwise Wilcoxon signed-rank tests, resulting in a significance level set at p≤0.001 indicated execution time differences between the following pairs only (Table 6). Therefore, we conclude that any learning rate value between [0.004, 0.010] offers the best results, but only in terms of execution time.

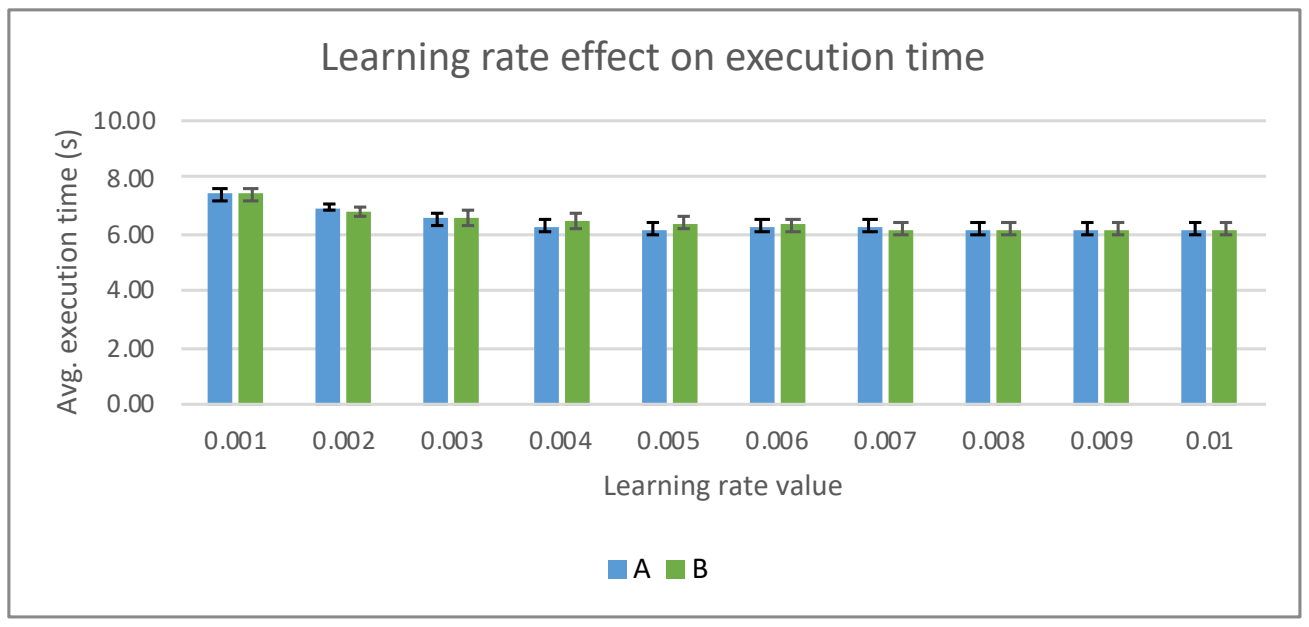

**Figure 4. Effect of learning rate value parameter on NN execution time (momentum=0.3). Error bars at 95%c.i.**

| | Execution time (s) | | Test statistics | |
|---|---|---|---|---|
| **Momentum value pairs** | **Mean** | **Std. Dev.** | **Z** | **p** |
| 0.8 - 0.3 | 5.20 - 6.40 | 0.40 - 0.49 | -3.286 | 0.001 |
| 0.8 - 0.4 | 5.20 - 6.20 | 0.40 - 0.40 | -2.236 | 0.001 |
| 0.8 - 0.5 | 5.20 - 6.20 | 0.40 - 0.40 | -3.419 | 0.001 |
| 0.8 - 0.6 | 5.20 - 5.87 | 0.40 - 0.34 | -3.612 | 0.001 |
| 0.8 - 0.7 | 5.20 - 5.87 | 0.40 - 0.34 | -3.612 | 0.001 |

**Table 6. Exp B. statistically significant differences of NN execution time for momentum values**

Proceeding with the same process for experiment B, we note that although nominal differences in the average accuracy emerge for values 0.009 and 0.010 (Figure 3), a Friedman test does not reveal a statistically significant difference between the accuracy at each learning rate setting ($\chi^2_{(9)}$=9.000, p=0.437), thus we did not proceed with pairwise comparisons. However, in terms of execution time (Figure 4), a Friedman test does reveal a statistically significant difference between the accuracy at each learning rate setting ($\chi^2_{(9)}$=79.859, p=0.000). Post-hoc Bonferroni corrected pairwise Wilcoxon signed-rank tests, resulting in a significance level set at p≤0.001 indicated execution time differences between the pairs shown in Table 7. Therefore we conclude that any learning rate value between [0.004, 0.010] offers the best results, but only in terms of execution time.

| | Execution time (s) | | Test statistics | |
|---|---|---|---|---|
| **LR value** | **Mean** | **Std. Dev.** | **Z** | **p** |
| 0.001 - 0.004 | 7.40 - 6.47 | 0.49 - 0.50 | -3.276 | 0.001 |
| 0.001 - 0.005 | 7.40 - 6.40 | 0.49 - 0.49 | -3.419 | 0.001 |
| 0.001 - 0.006 | 7.40 - 6.33 | 0.49 - 0.47 | -3.557 | 0.000 |
| 0.001 - 0.007 | 7.40 - 6.20 | 0.49 - 0.40 | -3.448 | 0.001 |
| 0.001 - 0.008 | 7.40 - 6.20 | 0.49 - 0.40 | -3.448 | 0.001 |
| 0.001 - 0.009 | 7.40 - 6.20 | 0.49 - 0.40 | -3.448 | 0.001 |
| 0.001 - 0.010 | 7.40 - 6.20 | 0.49 - 0.40 | -3.448 | 0.001 |

**Table 7. Exp B. statistically significant differences of NN execution time for learning rate values**

**Using Support Vector Machines**

Parallel to the experiments using neural network classifiers, we performed another round of classification experiments using SVMs. For this, an SVM classifier was ran on the same 15 training & validation sets used in the NN experimentation. The most important parameter in SVMs is the type of kernel used, and for this purpose we experimented with 6 different kernel types (dot, radial, polynomial, neural, ANOVA and multiquadric). For each kernel various options can be set, however we selected the default recommendations as set in RapidMiner studio.

*Experiments A&B – accuracy*

In terms of classification accuracy, we note a disparity of performance in the different kernels used during experiment A, with the multiquadtric kernel offering the best nominal average accuracy (m=86.67%, stdev=33.99%) (Figure 5). Notably however, the multiquadric kernel has offered a 100% classification accuracy in all but two cases, thereby we can only attribute this performance to the nature of those two datasets. A Friedman test for experiment A reveals a statistically significant difference between the accuracy with each kernel ($\chi^2_{(5)}$=41.679, p=0.000). Post-hoc Bonferroni corrected pairwise Wilcoxon signed-rank tests, resulting in a significance level set at p≤0.0033 indicated accuracy differences only between the neural and polynomial kernels (Z=-2.954, p=0.003).

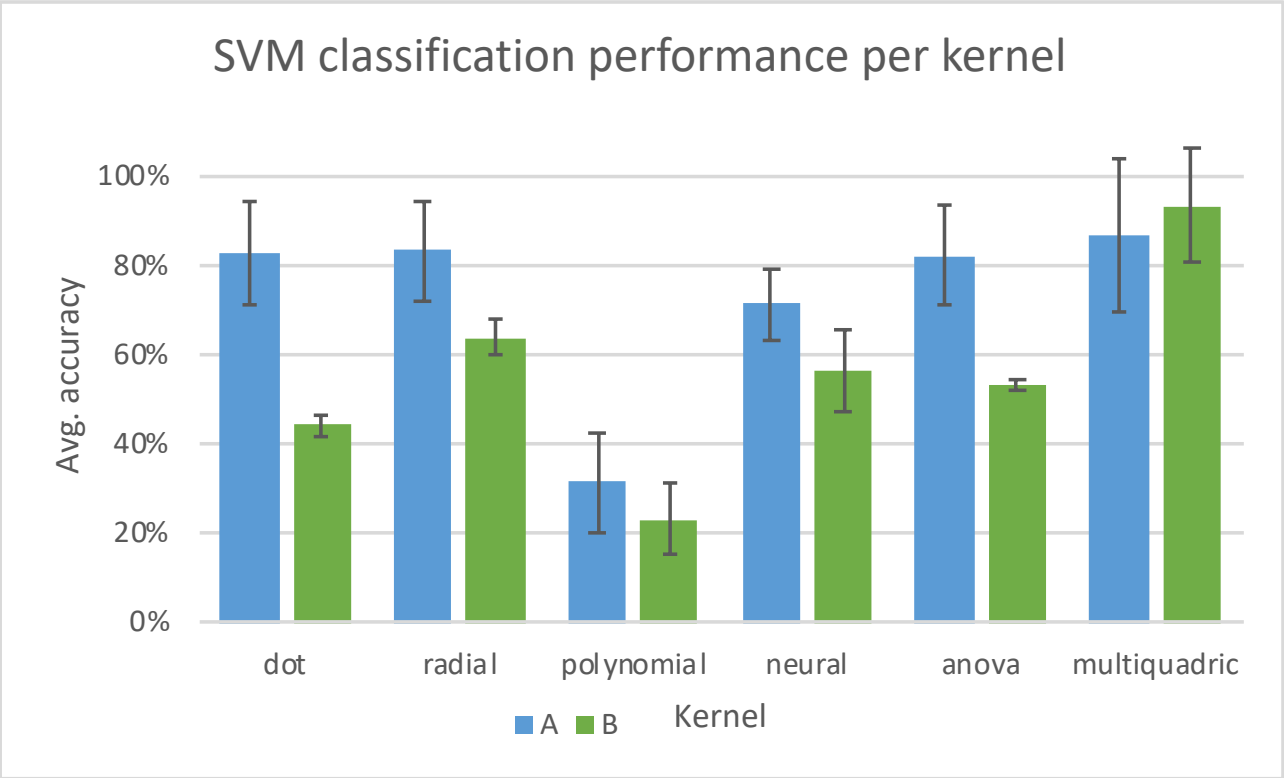

**Figure 5. SVM kernel accuracy. Error bars at 95%c.i.**

For experiment B, a disparity of performance emerges again, but this time the multiquadric kernel appears to have a more solid advantage, offering again the best performance (m=93.33%, stdev=24.94%) (Figure 5). Notably, again, the multiquadric kernel has offered a 100% accuracy on all but one cases. A Friedman test reveals a statistically significant difference between the accuracy with each kernel ($\chi^2_{(5)}$=49.933, p=0.000). Post-hoc Bonferroni corrected pairwise Wilcoxon signed-rank tests, resulting in a significance level set at p≤0.0033 indicated statistically significant execution time differences between the following pairs (Table 8).

| | Accuracy | | Test statistics | |
|---|---|---|---|---|
| **Kernel pairs** | **Mean** | **Std. Dev.** | **Z** | **p** |
| Radial - Dot | 63.88% - 44.08% | 7.58% - 4.72% | -3.126 | 0.002 |
| Polynomial - Dot | 23.17% - 44.08% | 15.47% - 4.72% | -3.238 | 0.001 |
| Anova - Dot | 53.25% - 44.08% | 2.33% - 4.72% | -3.098 | 0.002 |
| Polynomial - Radial | 23.17% - 63.88% | 15.47% - 7.58% | -3.294 | 0.001 |
| Neural - Polynomial | 56.41% - 23.17% | 18.00% - 15.47% | -3.181 | 0.001 |
| Anova - Polynomial | 53.25% - 23.17% | 2.33% - 15.47% | -3.238 | 0.001 |

**Table 8. Exp B. statistically significant differences of SVM accuracy for different kernels**

*Experiments A&B – execution time*

In terms of execution time, for Experiment A, different kernels seem to result in variability in execution time (Figure 6), although notably overall execution is very quick (<2s). A Friedman test for experiment A reveals a statistically significant difference between the execution time with each kernel ($\chi^2_{(5)}$=57.087, p=0.000). Post-hoc Bonferroni corrected pairwise Wilcoxon signed-rank tests, resulting in a significance level set at p≤0.0033 indicated execution time differences between the following pairs only (Table 9). From these results it appears that the multiquadric kernel, which offered the best precision, is also as fast to compute as simpler and less well-performing kernels (i.e. dot, polynomial, anova).

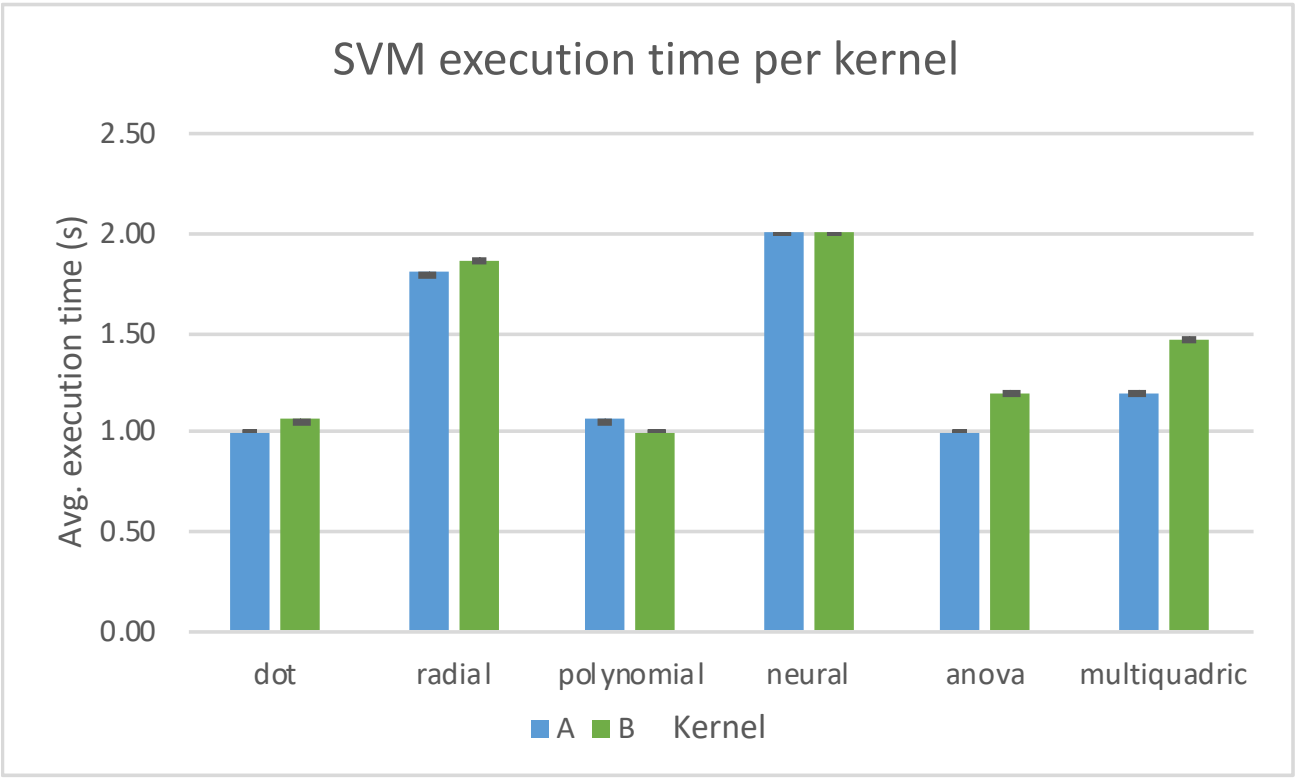

**Figure 6. SVM kernel execution time. Error bars at 95%c.i.**

| | Execution time (s) | | Test Statistics | |
|---|---|---|---|---|
| **Kernel pairs** | **Mean** | **Std. Dev.** | **Z** | **p** |
| Radial - Dot | 1.80 - 1.00 | 0.40 - 0.00 | -3.464 | 0.001 |
| Neural - Dot | 2.00 - 1.00 | 0.00 - 0.00 | -3.873 | 0.000 |
| Polynomial - Radial | 1.07 - 1.80 | 0.25 - 0.40 | -3.317 | 0.001 |
| Anova - Radial | 1.00 - 1.80 | 0.00 - 0.40 | -3.464 | 0.001 |
| Multiquadric - Radial | 1.20 - 1.80 | 0.40 - 0.40 | -3.000 | 0.003 |
| Neural - Polynomial | 2.00 - 1.07 | 0.00 - 0.25 | -3.742 | 0.000 |
| Anova - Neural | 1.00 - 2.00 | 0.00 - 0.00 | -3.873 | 0.000 |
| Multiquadric - Neural | 1.20 - 2.00 | 0.40 - 0.00 | -3.464 | 0.001 |

**Table 9. Exp A. statistically significant differences of SVM execution time for different kernels**

For Experiment B (Figure 6), a Friedman test reveals a statistically significant difference between the execution time with each kernel ($\chi^2_{(5)}$=49.463, p=0.000). Post-hoc Bonferroni corrected pairwise Wilcoxon signed-rank tests, resulting in a significance level set at $p \leq 0.0033$ indicated execution time differences between the following pairs only (Table 10). From these results it appears that the multiquadric kernel, which offered the best precision, is also as fast to compute as simpler and less well-performing kernels (i.e. dot, polynomial, anova).

| | Execution time (s) | | Test Statistics | |
|---|---|---|---|---|
| **Kernel pairs** | **Mean** | **Std. Dev.** | **Z** | **p** |
| Radial - Dot | 1.87 - 1.07 | 0.34 - 0.25 | -3.207 | 0.001 |
| Neural - Dot | 2.00 - 1.07 | 0.00 - 0.25 | -3.742 | 0.000 |
| Polynomial - Radial | 1.00 - 1.87 | 0.00 - 0.34 | -3.606 | 0.000 |
| Neural - Polynomial | 2.00 - 1.00 | 0.00 - 0.00 | -3.873 | 0.000 |
| Anova - Neural | 1.20 - 2.00 | 0.40 - 0.00 | -3.464 | 0.001 |

**Table 10. Exp B. statistically significant differences of SVM execution time for different kernels**

*Reverse procedure*
Intrigued by the surprisingly good performance of the SVM classifier, we repeated the process but this time in reverse, i.e. training the SVM classifier with the data of the validation sets (DS1-F & DS2-F) and classifying the contents of the 15 original mixed training sets, assessing the classifier's ability to correctly identify the sessions which were derived from the field. In this case, we noted a 100% (sd=0.00%) accuracy with all the 30 datasets in both experiments A & B. This is an even more astounding result, because it would be logical to assume that the classifier should perform better when trained with samples belonging to both prediction outcomes. However, it seems that the inherent differences in entry behaviour, as captured by and modeled using our four attributes, are enough that a classifier can be adequately trained using samples from just one category. For reference, the multiquadric kernel settings used were kernel sigma1 = 1.0, kernel shift =1.0, kernel cache=200, c=0.0, convergence e = 0.001, max iterations=100000, scale=true, L pos=1.0, L neg=1.0, epsilon=0.0, epsilon plus=0.0, epsilon minus=0.00, balance cost= false, quadratic loss pos=false, quadratic loss neg=false.

**The effect of each variable in classification results**

From the previous process, we note that the best possible results were obtained using neural networks, where we were able to achieve 100% classification accuracy with a range of momentum and learning rate parameters. Typically, neural network behaviour is seen as a "black box" process that cannot lead to the examination of effects of input parameters. However, an examination of the produced weights between neurons can yield useful insights (Olden & Jackson, 2002). Such a method was proposed in (Garson, 1991) and has been studied in several application cases, e.g. (Goh, 1995), (Gevrey, Dimopoulos, & Lek, 2003).

In our previous analysis, our purpose was to find the best possible accuracy that the NN could provide, hence we used a large number of training cycles (10,000) which resulted in execution times spanning several seconds. Since we were able to determine a range of momentum and learning rate parameters in which optimal performance could be gained, we aimed to reduce the number of training cycles in which this could be achieved. Using one of our subsets for experiments A & B we performed a hyperparameter optimization process using evolutionary algorithms. These algorithms can produce results that are close to optimal and have the advantage of converging reasonably quickly, compared to exhaustive grid searches. For each subset, we set appropriate thresholds for the hyperparameters (momentum [0.01, 1.0], learning rate [0.001, 0.01], training cycles [50, 1000] and ran the evolutionary algorithm 25 times using a random seed, and recorded the calculated recommendations. For Experiment A, we noted that the minimum number of training cycles in which a perfect accuracy could be achieved was 223 (max = 983), while in Experiment B, it was 163 (max=998). Therefore, we decided to empirically select a value of 200 training cycles, as a figure between the minima observed for the two experiments. For this number of training cycles, we repeated the process 25 more times for each experiment. In each run the optimizer was able to find a 100% accuracy solution, and we recorded the produced LR and momentum values (see Table 11), which we note are quite close for both experiments.

| | Momentum | | Learning rate | |
|---|---|---|---|---|
| **Experiment** | **Mean** | **Std. Dev.** | **Mean** | **Std. Dev.** |
| A (n=25) | 0.513 | 0.243 | 0.006 | 0.003 |
| B (n=25) | 0.547 | 0.154 | 0.004 | 0.003 |

**Table 11. Average hyperparameter values yielding 100% accuracy in 200 training cycles (sample set 1)**

Using these parameters, we repeated the entire process using all data subsets, as described in the previous sections. For experiment A, the average accuracy achieved was 86.7% (sd=35%), effectively caused by having 13/15 datasets with 100% accuracy and 0% accuracy in two datasets. For these 13 datasets where accuracy was optimal, vertical finger slippage (Y-diff) (m=27.1%, sd=0.05%) and interkey time (27.2%, sd=0.07%) showed the greatest relative importance to the outcome variable showing predicted lab or field classification (Figure 7). For experiment B, the average accuracy was 93.3% (sd=25.8%), in this case 14/15 datasets showing 100% accuracy and 0% accuracy in one dataset. In this case, Garson's algorithm showed that keypress duration had the greatest relative importance (28.8%, sd=9.6%), followed by vertical finger slippage again (25.4%, sd=8.1%) (Figure 7).

In both cases, we note that the difference across variables seem quite small. For both experiments, pairwise t-tests comparing the mean relative importance the variables (with Bonferroni correction resulting in a significance level of p=0.0083) showed that the differences are not statistically significant (Table 12). Similarly, we were not able to detect a statistically significant difference in the relative importance of variables when comparing across experiments A and B (Table 13). In summary, these results mean that based on Garson's algorithm, all variables were equally important in achieving the observed classification results, meaning that we cannot attribute differences in field and lab text entry behaviour to specific variables, as they all seem to play an equal part.

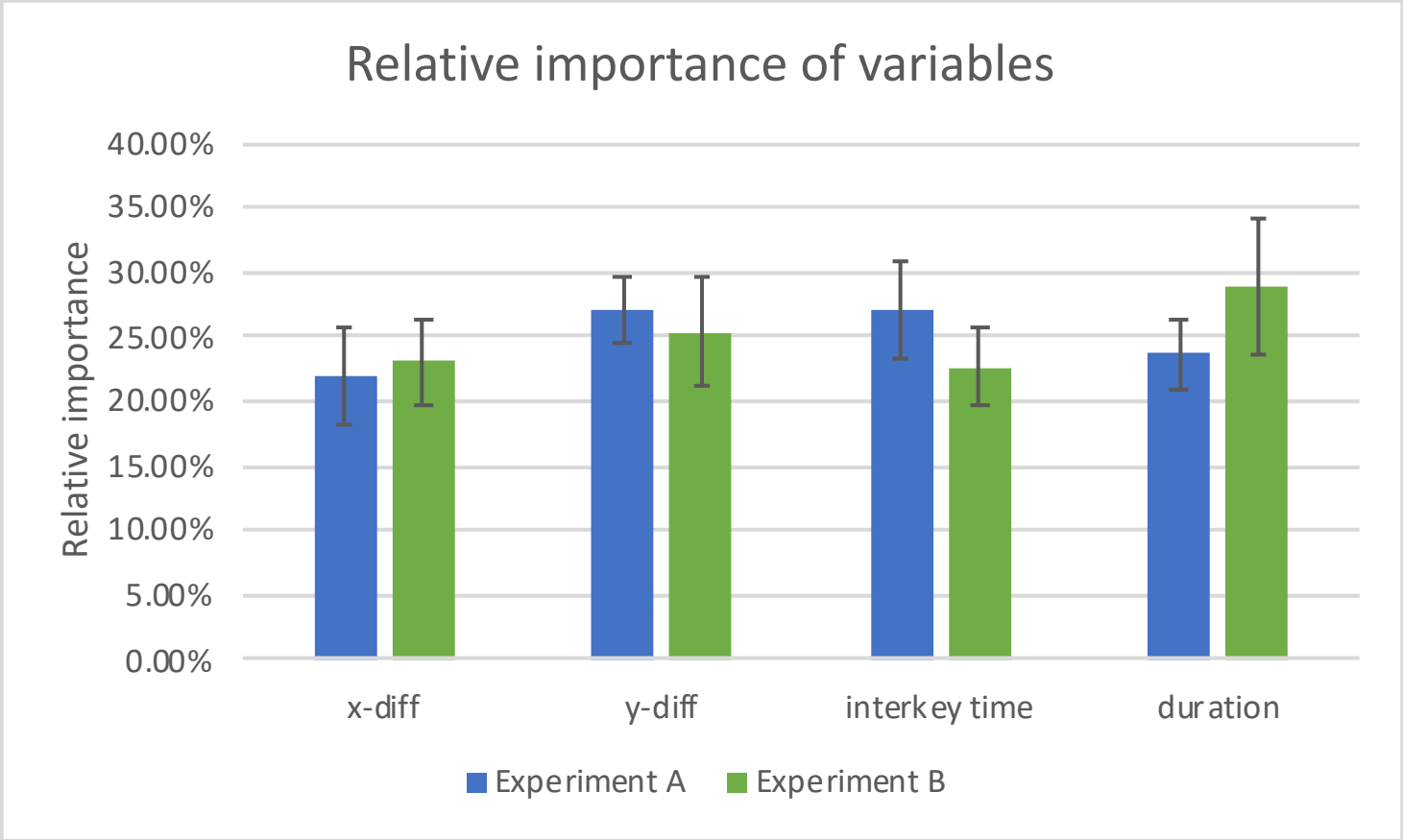

**Figure 7. Relative importance (Garson's algorithm) of NN input variables. Error bars at 95%c.i.**

| | | Experiment | T | Df | p (2-tailed) |
|---|---|---|---|---|---|
| A | Pair 1 | xdiff - ydiff | -2.143 | 12 | .053 |
| | Pair 2 | xdiff - interkey | -1.466 | 12 | .168 |
| | Pair 3 | xdiff - duration | -.542 | 12 | .598 |
| | Pair 4 | ydiff - interkey | -.014 | 12 | .989 |
| | Pair 5 | ydiff - duration | 1.641 | 12 | .127 |
| | Pair 6 | interkey - duration | 1.482 | 12 | .164 |
| B | Pair 1 | xdiff - ydiff | -.820 | 13 | .427 |
| | Pair 2 | xdiff - interkey | .166 | 13 | .871 |
| | Pair 3 | xdiff - duration | -1.594 | 13 | .135 |
| | Pair 4 | ydiff - interkey | .990 | 13 | .340 |
| | Pair 5 | ydiff - duration | -.802 | 13 | .437 |
| | Pair 6 | interkey - duration | -1.833 | 13 | .090 |

**Table 12. Pairwise T-test results for variable relative importance**

| | Eq. Var. assumed | t | df | Sig. (2-tailed) |
|---|---|---|---|---|
| xdiff | Yes | -0.396 | 25 | 0.695 |
| | No | -0.395 | 23.994 | 0.697 |
| ydiff | Yes | 0.664 | 25 | 0.512 |
| | No | 0.676 | 21.852 | 0.506 |
| interkey | Yes | 1.868 | 25 | 0.074 |
| | No | 1.85 | 22.614 | 0.077 |
| duration | Yes | -1.733 | 25 | 0.095 |
| | No | -1.773 | 19.581 | 0.092 |

**Table 13. Independent-sample T-test results for variable relative importance**

## DISCUSSION & CONCLUSIONS

In this paper, we set out to investigate whether text entry behaviour differs in the lab settings and the real world. Traditionally, input speed (WPM) and error rates are the focus of text entry studies, but both these metrics are composite results of a number of factors, such as the shifting of attention focus between the keyboard and text entry area, or between the device and external stimuli from the world, as well as factors like the way devices are held, the mobility of users and reflection taking place while typing. Many of these aspects are simply not accounted for in text entry studies. Because these mostly take place in lab environments, most of these factors are controlled for, therefore a direct comparison of speed and error rates between lab and field settings is inappropriate, or at the very least, provides no real explanation for the cause of disparities in performance. For these reasons, we set out to compare lab and field text entry behaviour on metrics that are more directly comparable, such as finger slippage, interkey time and duration of keypresses.

Furthermore, because of the data available for this study, we selected to proceed with a machine learning approach instead of comparing data with a traditional statistical approach. The results from our analyses were beyond our expectations. We were able to achieve almost perfect accuracy in correctly classifying the text entry behaviour of a population based on the data of two other independent populations, using both NN and SVM techniques and for the latter, even when training with samples from field use only. In doing so, we uncover evidence that text entry behaviour differs between lab and real-world use, and that finger slippage, inter-key time and keypress duration are predictors of this behavioral difference.

Machine learning classifiers, no matter their accuracy, are often considered to offer at best a "black box" view of the problem domains in which they are applied. In contrast, using inferential statistics, we are usually able to attribute observed differences to the effects of specific variables. Similar to the explanatory ability of inferential statistics, analysis of derived weights in NN classification models can indicate the relative importance of input variables on the outcome. Using Garson's algorithm (Garson, 1991), we found evidence that all the text entry behaviour variables that we examined play an equal role in achieving the high classification performance that we observed. Therefore, we cannot say that lab and field text entry behaviour differs solely because of finger slippage, or touch behaviour discrepancies, but that all these factors attribute to the observed behavioral differences. Of course, the precise mechanism with which each variable contributes remains the subject of future work. In this sense, our results seem akin to many of the studies in the medical domain which link substances or behaviours with health outcomes, even though the mechanism is not explained (e.g. think of the early tobacco-cancer relationship studies). Such studies can have profound impacts in the scientific community, by disrupting established knowledge and showing where future research might focus.

Here, we believe, lies our primary contribution for this paper. Namely, to show that text entry in the lab and in the real world differs, and therefore to argue that we need to start complementing our work in the lab with more evaluations in field settings. While for some this conclusion might be intuitive, to our knowledge, this is the first paper to offer tangible evidence for this view. Indeed, since the ability to develop full replacement IMEs has been around since 2011, and, given the fact that until now hardly any field study of mobile text entry has been carried out, it seems that for the scientific community, the fact that text entry behaviour in the lab is different from the field, seems not to be such an intuitive conclusion after all. Given the continued focus of the text entry research community on lab studies, despite the admittedly readily accessible technical capacity to evaluate text entry in the wild, we hope that the provision of this tangible evidence serves to spark a practice shift and encourage more researchers to begin exploring real-world text entry.

### Implications for field studies and design for real-world use of smartphone text entry methods

Real world use involves a variety of bodily posture contextual settings, where the user might be standing, seated, mobile, inputting with one or two hands etc. In our case, can we find that a difference between lab and field study text entry behaviour seems to exist. Our results indicate that this is not due to hand or bodily posture only – presumably our field participants often entered text in comfortable and quiet environments too, so if this were the only confounding factor, we would expect to at least see some false positives or negatives during classification. We believe that the nature of these differences must be intrinsic to how people use their devices for text entry in the real-world, and that future research should place more focus on the cognitive and psychophysical context of use, rather than attempt to focus on physical posture during use in field study data collection. Such factors can likely explain the contribution of interkey and keypress duration times in the classification outcomes, as we will discuss below. In this regard, we believe that the second contribution of our paper is to demonstrate the possibilities that can be opened up by field research in text entry. These stem from our consideration of finger slippage as a precursor to erroneous input and non-Fitts time metrics relating to cognitive processing during input.

Firstly, because of the finger slippage, we note the necessity of expanding existing work on touch models by collecting finger slippage data from field studies. In works such as (Azenkot & Zhai, 2012), (Tani & Yamada, 2015) and (Weir et al., 2014), touch data collected from lab sessions have shown promise in improving text entry performance, but this finding has only been tested in the lab. Based on our findings, the intuitive assumption that if such touch models are to be tested in the field, then they need to be built upon touch data also collected in field settings, becomes compelling.

Secondly, we note that since inter-key times and keypress durations are also predictors for text entry situatedness, more work is needed in understanding why this phenomenon emerges and how keyboards can offer better support in real world settings. We are primarily here concerned with how inter-key time can affect real world input speed (e.g. especially if it is caused by thinking what to enter next, reviewing typed text for possible errors, or by being an indicator that the user's attention has shifted away from the keyboard and towards the input area or the surrounding environment). Finding new ways to support the cognitive burden during real-world text entry is a promising direction towards increasing participants' performance with keyboards. Additionally, we are concerned with how keypress duration also impacts entry speed and possibly, combined with finger slippage, the error rates in touchscreen keyboards. Given the preliminary evidence linking increased keypress duration to stress by (Ciman, Wac, & Gaggi, 2015), we believe that adaptive levels of text entry support (e.g. more aggressive autocorrection, touch area shifting, input recommendation etc) might be beneficial to assist in situations where entry under stressful situations is detected.

### Implications for lab-based studies in mobile text entry

Our findings apply to the improvement of future lab-based studies as well, and in this lies the third contribution of our paper. In knowing that user text entry behaviour differs in the typical transcription-based text entry task and the real world, we are compelled to ask if the community should invest more effort away from novel keyboard designs or algorithmic support and move towards devising new and better methods for assessing text entry in the lab. The advantages of lab studies in terms of the invested time and internal validity are not going to disappear and, indeed, our view is that we absolutely need good lab studies before moving to assess performance in the field. However, it's becoming clear that we need to question the utility of the transcription task, in terms of its ability to approximate real-world use. So far only two studies proposing an alternative type of task exists (Vertanen & Kristensson, 2014)(Nicol, Komninos, & Dunlop, 2016), but even in these evaluations, the authors do not explore cognition or attention factors.

Firstly thus, our proposal to use low-level metrics which characterize text entry behaviour is a good first step towards knowing how to measure the extent to which any lab-based method adequately approximates real use, and it might well be the case that more such low-level metrics can be devised by the community. For this we also need larger datasets of real-world text entry behaviour, but this is not an unfeasible task if the community decide to engage in more field studies and share data more openly.

Secondly, we need to consider how we might improve the transcription task, if we are to keep it as a part of our text entry evaluation toolkit. So far, it has been assumed that user attention is not diverted during text entry studies, but in reality, users perform visual checks on the text they entered and this may have a slowing-down effect (Clawson, Lyons, Starner, & Clarkson, 2005). The extent to which this phenomenon applies to performance during transcription tasks (especially for novel input methods) that are the de-facto task type in lab studies remains largely unexplored and uncontrolled for. More research is needed here, possibly with the assistance of eye-tracking or neurosignal tools.

Finally, keypress duration in both our field datasets was also found to be lower on average compared to the lab study. Since there exists preliminary evidence linking increased keypress duration to stress (Ciman et al., 2015), further exploration on the stress levels of lab study participants needs to take place, in order to assess whether this increased stress level is a product of intense focus on the task (under knowledge of being monitored) and thus artificially impacting both entry speed and error rates. Associated with this, we believe that further investigation must be made on the phenomenon of improved participant performance over repeated text entry sessions in the lab. We are especially concerned with how much of the observed gains are due to practice and familiarity with the input method, or in fact, with the process of being a lab study participant.

## ACKNOWLEDGEMENTS AND REPLICABILITY INFORMATION

This work is partly funded by project [PROJECT IDENTIFIER LEFT BLANK FOR REVIEW]. The data used in this study as well as RapidMiner Studio project files to allow the replication of results are available publicly at [WEBSITE URL LEFT BLANK FOR REVIEW].